\documentclass[preprintnumbers,amsmath,amssymbm,prd]{revtex4}
\usepackage{epsfig}
\usepackage{graphicx}

\begin{document}
\title{Super-extremal black holes in the quasitopological electromagnetic field theory}
\author{Shahar Hod}
\address{The Ruppin Academic Center, Emeq Hefer 40250, Israel}
\address{ }
\address{The Hadassah Institute, Jerusalem 91010, Israel}
\date{\today}

\begin{abstract}
\ \ \ It has recently been proved that a simple generalization of
electromagnetism, referred to as quasitopological electromagnetic field theory, 
is characterized by the presence of dyonic black-hole solutions of the Einstein field equations that, 
in certain parameter regions, are characterized by four horizons. 
In the present compact paper we reveal the 
existence, in this non-linear electrodynamic field theory, 
of super-extremal black-hole spacetimes that are characterized by the four degenerate functional 
relations $[g_{00}(r)]_{r=r_{\text{H}}}=[dg_{00}(r)/dr]_{r=r_{\text{H}}}=[d^2g_{00}(r)/dr^2]_
{r=r_{\text{H}}}=[d^3g_{00}(r)/dr^3]_{r=r_{\text{H}}}=0$, where $g_{00}(r)$ is the $tt$-component 
of the curved line element and $r_{\text{H}}$ is the black-hole horizon radius. 
In particular, using analytical techniques we prove that the quartically degenerate super-extremal black holes are characterized by the 
universal (parameter-{\it independent}) dimensionless compactness parameter 
$M/r_{\text{H}}={2\over3}(2\gamma+1)$, 
where $\gamma\equiv{_2F_1}(1/4,1;5/4;-3)$.
\end{abstract}
\bigskip
\maketitle


\section{Introduction}

The canonical Reissner-Nordstr\"om solution \cite{Chan} of the coupled Einstein-Maxwell field equations describes 
a two-parameter family of asymptotically flat black-hole spacetimes. 
These curved spacetimes are characterized by two conserved (asymptotically measured) physical parameters: 
the mass $M$ and the electric charge $Q$ of the central black hole. 

In standard Maxwell electrodynamics the boundary between black-hole spacetimes and horizonless 
naked singularities is marked by the presence of a zero-temperature extremal black hole 
whose horizon radius $r=r_{\text{H}}$ is characterized by the doubly degenerate functional relations
\begin{equation}\label{Eq1}
g_{00}(r)={{dg_{00}(r)}\over{dr}}=0\ \ \ \ \ \text{for}\ \ \ \ \ r=r_{\text{H}}\  ,
\end{equation}
where the radially-dependent metric function of the curved spacetime is given by the expression 
$g_{00}(r)=1-2M/r+Q^2/r^2$. From (\ref{Eq1}) one finds that the 
extremal Reissner-Nordstr\"om black hole is characterized by the familiar dimensionless relations \cite{Noteunits} 
\begin{equation}\label{Eq2}
{{|Q|}\over{M}}={{r_{\text{H}}}\over{M}}=1\  .
\end{equation}

Recently, a physically intriguing (and mathematically elegant) generalization of Maxwell's 
electromagnetism, referred to as quasitopological electromagnetic field theory, has 
been studied in \cite{qua1} (see also \cite{qua2,qua3,qua4} and references therein). 
As discussed in \cite{qua1}, since the effects of the generalized non-linear 
quasitopological electromagnetic field theory are not manifest in Earth-based experiments, 
it may provide a reasonable phenomenologically rich alternative to the standard Maxwell field theory. 

Intriguingly, it has been shown in \cite{qua1} that, when coupled to gravity, 
the newly suggested quasitopological electromagnetic field theory is characterized by the presence of asymptotically flat dyonic black-hole spacetimes that, as opposed to the standard Reissner-Nordstr\"om black-hole solutions 
of the Einstein-Maxwell theory, may have four horizons. 

The main goal of the present compact paper is to reveal the physically interesting fact that 
the non-linear electrodynamic field theory \cite{qua1} is characterized by the presence of a unique family of 
super-extremal black-hole spacetimes whose horizons are {\it quartically} degenerate. 
In particular, below we shall use analytical techniques in order to determine the 
physical parameters that characterize these unique super-extremal black holes of the non-trivial (non-linear) 
electromagnetic field theory. 

\section{Super-extremal black holes in the quasitopological non-linear electromagnetic field theory}

The quasitopological electromagnetic field theory is characterized by the composed action \cite{qua1}
\begin{equation}\label{Eq3}
S={{1}\over{16\pi}}\int{\sqrt{-g}d^4x\{R-\alpha_1 F^2-\alpha_2[(F^2)^2-2F^{(4)}]\}}\  ,
\end{equation}
where $F^2=F^{\mu\nu}F_{\mu\nu}$, 
$F^{(4)}=F^{\mu}_{\nu}F^{\nu}_{\rho}F^{\rho}_{\sigma}F^{\sigma}_{\mu}$, 
and $\{\alpha_1,\alpha_2\}$ are the coupling parameters of the non-linear field theory \cite{Notey12}. 
Note that the standard Maxwell theory is characterized by the simple relations $\alpha_1=1$ and 
$\alpha_2=0$. 

Assuming a spherically symmetric curved spacetime of the form \cite{Chan,Notesch}
\begin{equation}\label{Eq4}
ds^2=-g_{00}(r)dt^2+g_{11}(r)dr^2+r^2(d\theta^2+\sin^2\theta d\phi^2)\  ,
\end{equation}
one finds that the quasitopological electromagnetic field theory (\ref{Eq3}), when coupled to the 
Einstein equations, is characterized by the presence of dyonic black-hole solutions of mass $M$, 
electric charge $q$, and magnetic charge $p/\alpha_1$ of the form \cite{qua1}
\begin{equation}\label{Eq5}
g_{00}(r)=[g_{11}(r)]^{-1}=1-{{2M}\over{r}}+{{\alpha_1 p^2}\over{r^2}}+{{q^2}\over{\alpha_1 r^2}}\cdot
{_2F_1}\Big({1\over4},1;{5\over4};-{{4p^2\alpha_2}\over{\alpha_1 r^4}}\Big)\  ,
\end{equation}
where ${_2F_1}(a,b;c;z)$ is the hypergeometric function \cite{Abram}. 
As shown in \cite{qua1}, for this non-vacuum spacetime, the null energy condition, the weak energy 
condition, and the dominant energy condition are all respected in the 
regime $\alpha_1>0$ with $\alpha_2>0$, whereas the strong energy condition is violated. 

Before proceeding, it is worth stressing the fact that the presence of the hypergeometric function ${_2F_1}(a,b;c;z)$ 
in the curved line element (\ref{Eq5}) makes it a highly non-trivial task to explore, 
using purely analytical techniques, 
the physical and mathematical properties of these dyonic black-hole spacetimes. 
Nevertheless, we shall now prove explicitly that some interesting features 
of the recently proposed non-linear electrodynamic field theory (\ref{Eq3}) 
can be deduced from a direct inspection of the non-trivial curved line element (\ref{Eq5}). 

In particular, we shall prove that the 
quasitopological electromagnetic field theory (\ref{Eq3}) is characterized by the presence 
of super-extremal black-hole spacetimes which are characterized by the {\it quartically} 
degenerate functional relations 
\begin{equation}\label{Eq6}
g_{00}(r)={{dg_{00}(r)}\over{dr}}={{d^2g_{00}(r)}\over{dr^2}}={{d^3g_{00}(r)}\over{dr^3}}=0
\ \ \ \ \ \text{for}\ \ \ \ \ r=r_{\text{H}}\  .
\end{equation}
[Compare (\ref{Eq6}) with the doubly degenerate relations (\ref{Eq1}) that characterize the extremal black-hole 
solutions of the standard Maxwell theory.] 
In particular, using analytical techniques we shall determine the physical parameters 
that characterize the super-extremal black holes of the quasitopological non-linear electromagnetic field theory (\ref{Eq3}). 

To this end, it is convenient to define the dimensionless physical parameters 
\begin{equation}\label{Eq7}
{\bar M}\equiv {{M}\over{r_{\text{H}}}}\ \ \ \ ; \ \ \ \ {\bar p}\equiv {{p}\over{r_{\text{H}}}}\ \ \ \ ; \ \ \ \ 
{\bar q}\equiv {{q}\over{r_{\text{H}}}}\ \ \ \ ; \ \ \ \ {\bar\alpha_2}\equiv {{\alpha_2}\over{r^2_{\text{H}}}}\
\end{equation}
of the black-hole spacetime. 
We shall also use the gradient relation \cite{Abram}
\begin{equation}\label{Eq8}
{{d\Big[{_2F_1}\Big({1\over4},1;{5\over4};-{{4p^2\alpha_2}\over{\alpha_1 r^4}}\Big)\Big]}\over{dr}}=
{{{_2F_1}\Big({1\over4},1;{5\over4};-{{4p^2\alpha_2}\over{\alpha_1 r^4}}\Big)}\over{r}}-
{{r^3}\over{r^4+{{4p^2\alpha_2}\over{\alpha_1}}}}\
\end{equation}
which characterizes the hypergeometric function.

We first note that, taking cognizance of Eqs. (\ref{Eq5}) (\ref{Eq7}), and (\ref{Eq8}), 
one finds that dyonic black holes with triply degenerate horizons 
[$g_{00}(r)={{dg_{00}(r)}/{dr}}={{d^2g_{00}(r)}/{dr^2}}=0$] 
are determined by the coupled functional relations
\begin{equation}\label{Eq9}
{2\bar M}=\alpha_1{\bar p}^2+{{\gamma{\bar q}^2}\over{\alpha_1}}+1\  ,
\end{equation}
\begin{equation}\label{Eq10}
{2\bar M}=2\alpha_1{\bar p}^2+{{[\gamma(c+1)+1]{\bar q}^2}\over{\alpha_1}(c+1)}\  ,
\end{equation}
and
\begin{equation}\label{Eq11}
{2\bar M}=3\alpha_1{\bar p}^2+{{[\gamma(c+1)^2+2]{\bar q}^2}\over{\alpha_1}(c+1)^2}\  .
\end{equation}
The (rather cumbersome) set of equations (\ref{Eq9}), (\ref{Eq10}), and (\ref{Eq11}) 
can be solved analytically to yield the dimensionless relations
\begin{equation}\label{Eq12}
{\bar M}={{\gamma(c+1)^2+3c-1}\over{4c}}\  ,
\end{equation}
\begin{equation}\label{Eq13}
{\bar q}=(c+1)\sqrt{{{\alpha_1}\over{2c}}}\  ,
\end{equation}
and
\begin{equation}\label{Eq14}
{\bar p}=\sqrt{{{c-1}\over{2c\alpha_1}}}\
\end{equation}
for dyonic black holes with triply degenerate horizons, 
where we have used here the dimensionless variables
\begin{equation}\label{Eq15}
c\equiv {{4p^2\alpha_2}\over{\alpha_1r^4_{\text{H}}}}={{4{\bar p}^2{\bar \alpha_2}}\over{\alpha_1}}>1\
\end{equation}
and
\begin{equation}\label{Eq16}
\gamma\equiv {_2F_1}\Big({1\over4},1;{5\over4};-c\Big)\  .
\end{equation}

Adding to Eqs. (\ref{Eq12}), (\ref{Eq13}), (\ref{Eq14}), (\ref{Eq15}), and (\ref{Eq16}) 
the fourth requirement, ${{d^3g_{00}(r)}/{dr^3}}=0$ [see Eq. (\ref{Eq6})], which can 
be expressed in the form
\begin{equation}\label{Eq17}
{6\bar M}=12\alpha_1{\bar p}^2+{{[3\gamma(c+1)^3+(c-3)^2]{\bar q}^2}\over{\alpha_1}(c+1)^3}\  ,
\end{equation}
one finds that the super-extremal (quartically degenerate) dyonic black holes 
are characterized by the remarkably compact dimensionless relations 
\begin{equation}\label{Eq18}
{\bar M}={2\over3}(2\gamma+1)\  ,
\end{equation}
\begin{equation}\label{Eq19}
{\bar q}=\sqrt{{{8\alpha_1}\over{3}}}\  ,
\end{equation}
\begin{equation}\label{Eq20}
{\bar p}={{1}\over\sqrt{3\alpha_1}}\  ,
\end{equation}
and \cite{Notecg2}
\begin{equation}\label{Eq21}
c=3\  .
\end{equation}
Taking cognizance of Eqs. (\ref{Eq16}) and (\ref{Eq21}), one obtains the relation
\begin{equation}\label{Eq22}
\gamma\equiv {_2F_1}\Big({1\over4},1;{5\over4};-3\Big)=
{{\ln\Big({{1+\sqrt{3}+\sqrt{2}\sqrt[4]{3}}\over{1+\sqrt{3}-\sqrt{2}\sqrt[4]{3}}}\Big)+
2\arctan\Big({{\sqrt{2}\sqrt[4]{3}}\over{2+\sqrt{2}\sqrt[4]{3}}}\Big)+
2\arctan\Big({{\sqrt{2}\sqrt[4]{3}}\over{2-\sqrt{2}\sqrt[4]{3}}}\Big)}\over
{4\sqrt{2}\sqrt[4]{3}}}\simeq0.746\  .
\end{equation}

Substituting Eqs. (\ref{Eq18}), (\ref{Eq19}), (\ref{Eq20}), (\ref{Eq21}), and (\ref{Eq22}) into Eq. (\ref{Eq5}), one 
finds the gradient relation
\begin{equation}\label{Eq23}
{{d^4g_{00}(r)}\over{dr^4}}={{12}\over{r^4_{\text{H}}}}>0
\ \ \ \ \ \text{for}\ \ \ \ \ r=r_{\text{H}}\  ,
\end{equation}
which implies that, for the super-extremal black holes, the quartically 
degenerate horizon $r=r_{\text{H}}$ is a minimum point 
of the metric function $g_{00}(r)$ [this fact is consistent with the asymptotic behaviors $g_{00}(r\to0)\to+\infty$ and 
$g_{00}(r\to\infty)\to1$, see Eq. (\ref{Eq5})]. 

\section{Summary}

Motivated by the physically interesting quasitopological non-linear electromagnetic field theory suggested 
in \cite{qua1} (see also \cite{qua2,qua3,qua4} and references therein), in the present compact paper 
we have studied the physical and mathematical properties of dyonic black-hole solutions of the theory. 
Intriguingly, we have revealed the existence 
of super-extremal black-hole spacetimes whose horizons are {\it quartically} degenerate [see Eq. (\ref{Eq6})]. 

In particular, using analytical techniques, 
we have proved that the quartically degenerate black holes are characterized by the 
dimensionless compactness parameter [see Eqs. (\ref{Eq7}), (\ref{Eq18}), and (\ref{Eq22})]
\begin{equation}\label{Eq24}
{{M}\over{r_{\text{H}}}}={2\over3}(2\gamma+1)\  .
\end{equation}
It is physically interesting to stress the fact that the 
dimensionless compactness ratio (\ref{Eq24}) of the super-extremal black holes 
is {\it universal} in the sense that it is independent of the coupling parameters $\{\alpha_1,\alpha_2\}$ that 
characterize the non-linear electromagnetic field theory (\ref{Eq3}). 

It is also worth pointing out that inspection of Eqs. (\ref{Eq2}), (\ref{Eq22}), and (\ref{Eq24}) reveals the fact 
that the super-extremal black holes of the quasitopological electromagnetic field theory (\ref{Eq3}) 
are more compact 
than the corresponding extremal Reissner-Nordstr\"om black holes of the standard Maxwell 
electromagnetic field theory. 

In addition, we have shown that the super-extremal 
dyonic black holes of the non-linear electromagnetic field theory (\ref{Eq3}) 
are characterized by the following dimensionless physical parameters [see 
Eqs. (\ref{Eq7}), (\ref{Eq15}), (\ref{Eq19}), (\ref{Eq20}), (\ref{Eq21}), (\ref{Eq22}), and (\ref{Eq24})] 
\begin{equation}\label{Eq25}
{{q}\over{M}}={{\sqrt{6\alpha_1}}\over{2\gamma+1}}\  ,
\end{equation}
\begin{equation}\label{Eq26}
{{p}\over{M}}={{\sqrt{3}}\over{2(2\gamma+1)\sqrt{\alpha_1}}}\  ,
\end{equation}
and
\begin{equation}\label{Eq27}
{{\alpha_2}\over{M^2}}=\Big[{{9\alpha_1}\over{4(2\gamma+1)}}\Big]^2\  .
\end{equation}

Finally, it is interesting to point out that the super-extremal black holes are defined by the 
set (\ref{Eq6}) of four equations that are expressed in terms of the 
six physical parameters $\{M,q,p,r_{\text{H}},\alpha_1,\alpha_2\}$. 
Thus, the physical parameters $\{M,q,p,r_{\text{H}}\}$ of the super-extremal dyonic black holes can be 
expressed solely in terms of the non-trivial coupling parameters $\{\alpha_1,\alpha_2\}$ of the composed field theory. 
In particular, using the dimensionless ratio (\ref{Eq27}), one finds the functional relations 
\begin{equation}\label{Eq28}
M={{4(2\gamma+1)}\over{9}}\cdot\sqrt{{{\alpha_2}}\over{\alpha^2_1}}\  ,
\end{equation}
\begin{equation}\label{Eq29}
q={{4\sqrt{2}}\over{3\sqrt{3}}}\cdot\sqrt{{{\alpha_2}}\over{\alpha_1}}\  ,
\end{equation}
\begin{equation}\label{Eq30}
p={{2}\over{3\sqrt{3}}}\cdot\sqrt{{{\alpha_2}}\over{\alpha^3_1}}\  ,
\end{equation}
and
\begin{equation}\label{Eq31}
r_{\text{H}}={{2}\over{3}}\cdot\sqrt{{{\alpha_2}}\over{\alpha^2_1}}\
\end{equation}
for the physical parameters of the super-extremal dyonic black holes that characterize the 
non-linear quasitopological electromagnetic field theory (\ref{Eq3}).

\bigskip
\noindent
{\bf ACKNOWLEDGMENTS}
\bigskip

This research is supported by the Carmel Science Foundation. I thank
Yael Oren, Arbel M. Ongo, Ayelet B. Lata, and Alona B. Tea for
stimulating discussions.


\begin{thebibliography}{99}

\bibitem{Chan} S. Chandrasekhar, {\it The Mathematical Theory of Black Holes}, (Oxford
University Press, New York, 1983).

\bibitem{Noteunits} We use natural units in which $G=c=\hbar=1$. 

\bibitem{qua1} H.-S. Liu, Z.-F. Mai, Y.-Z. Li, and H. Lü, 
Sci. China Phys. Mech. Astron. {\bf 63}, 240411 (2020) [arXiv:1907.10876]. 

\bibitem{qua2} Y.-Q. Lei, X.-H. Ge, and C. Ran, Phys. Rev. D {\bf 104}, 046020 (2021).

\bibitem{qua3} H. Huang, M.-Y. Ou, M.-Y. Lai, and H. Lü, Phys. Rev. D {\bf 105}, 104049 (2022).

\bibitem{qua4} S. W. Wei, Y. P. Zhang, Y. X. Liu, and R. B. Mann, Phys. Rev. Research {\bf 5}, 043050 (2023).

\bibitem{Notey12} Note that the coupling parameter $\alpha_1$ is dimensionless whereas 
the coupling parameter $\alpha_2$ has the dimensions of length$^2$.

\bibitem{Notesch} Here $\{t,r,\theta,\phi\}$ are the Schwarzschild coordinates of the 
spherically symmetric curved spacetime.

\bibitem{Abram} M. Abramowitz and I. A. Stegun, {\it Handbook of 
Mathematical Functions} (Dover Publications, New York, 1970).

\bibitem{Notecg2} Note that Eqs. (\ref{Eq15}), (\ref{Eq20}), and (\ref{Eq21}) imply the simple dimensionless relation 
${\bar \alpha_2}=(3\alpha_1/2)^2$. 


\end{thebibliography}
\end{document}